\documentclass[twocolumn]{article}
\usepackage{epsf}

\newcommand{\be}{\begin{equation}}
\newcommand{\ee}{\end{equation}}
\newcommand{\rd}{{\mathrm{d}}}
\newcommand{\rc}{{\mathrm{c}}}
\newcommand{\PS}{{\mathrm{PS}}}

\begin{document}

\title{{\bfseries Group Identification in N-Body Simulations: SKID and DENMAX Versus
  Friends-of-Friends}\footnote{Preprint BROWN HET-1140}}
\author{Martin G\"{o}tz\footnote{Harvard-Smithsonian Center
        for Astrophysics, 60 Garden St., MS-20, Cambridge,
        MA 02138} \footnote{Physics Department, Brown University,
        Box 1843, Providence, RI 02912} \footnote{To whom communication
        should be addressed. E-mail: {\ttfamily mgotz@cfa.harvard.edu}} ,
        John P. Huchra$^\dagger$,
        Robert H. Brandenberger$^\ddagger$
}
\date{\today}
\maketitle

\begin{abstract}
Three popular algorithms (FOF, DENMAX, and SKID) to identify halos in
cosmological N-body simulations are compared with each other and with
the predicted mass function from Press-Schechter theory. It is shown that
the resulting distribution of halo masses strongly depends upon the
choice of free parameters in the three algorithms, and therefore much
care in their choice is needed. For many parameter values, DENMAX and SKID
have the tendency to include in the halos particles at large distances
from the halo center with low peculiar velocities.
FOF does not suffer from this problem, and its mass distribution furthermore
is reproduced well by the prediction from Press-Schechter theory.
\end{abstract}

\section{Introduction}

N-body simulations are a frequently used tool in cosmology to study the
origin of large-scale structure in the universe or the formation of galaxies.
For these applications, it is necessary to identify the dark matter halos
in the models, as those are interpreted to be the locations of galaxies
or galaxy clusters. Ideally, the way in which the dark matter halos are
extracted should not influence the results sought in the simulations, e.g.\ 
the mass distribution, correlation function, formation and merging rates,
and many other properties of the halos. Fortunately or unfortunately, many
different algorithms to identify groups of particles in N-body simulations
have been proposed. The question we address here is how the choice of algorithm
affects the properties of halos.
\begin{figure*}[tb]
\epsfxsize=6 in \epsfbox{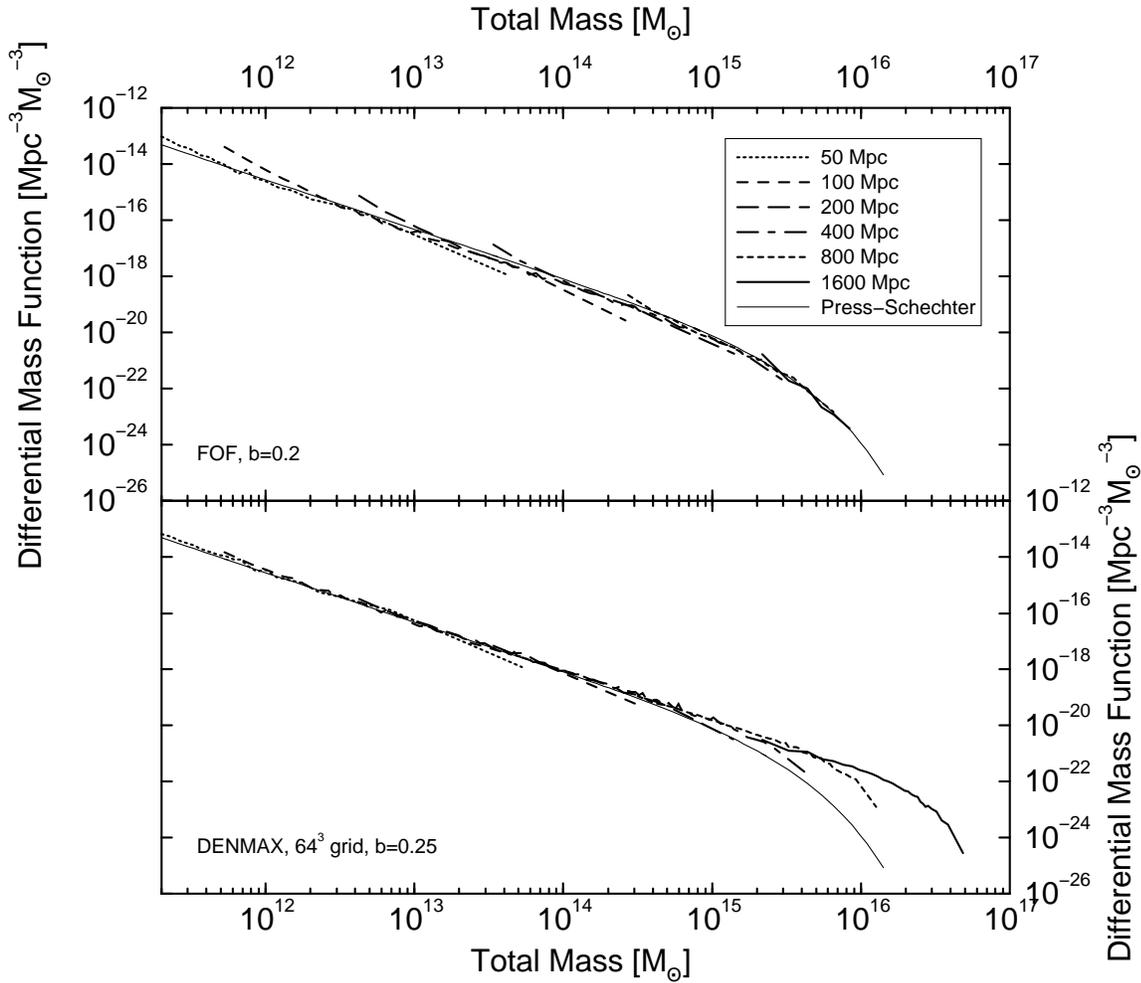}
\caption{The differential mass functions of halos for a series of
N-body simulations in an SCDM scenario ($\mathnormal{\Omega}_0=1$, $h=0.5$) at
redshift $z=0$, run with the P$^3$M code from \cite{bert}, for varying
box sizes, but with $64^3$ particles each. Groups were identified with
two popular algorithms: friends-of-friends (FOF) and DENMAX. Both have
as a free parameter a linking length between neighboring particles $b$,
here given in units of the mean particle separation. In addition, DENMAX
has as a second parameter the size of the grid on which the densities of
particles are calculated, chosen to be the same size as the number of
particles here. Compared to FOF or a theoretical prediction from
Press-Schechter theory, DENMAX produces more massive halos in large
simulations.}
\label{pfigure}
\end{figure*}

The attention of the authors was drawn to this question by the results shown
in Fig.\ \ref{pfigure}. This shows the mass distribution of halos, in
the form of the differential mass function, at redshift $z=0$ for a series
of N-body simulations, run with a particle-particle/particle-mesh (P$^3$M)
code by E. Bertschinger and J. M. Gelb \cite{bert}. The simulations,
originally carried out for a different purpose, are of a standard cold
dark matter (SCDM) model ($\mathnormal{\Omega}_0=1$ and $H_0=50 \mbox{ km/s/Mpc}$)
with $64^3$ particles in boxes of sizes varying from 50 Mpc to 1600 Mpc.
The mass distribution of halos was determined in two different ways, using
two popular algorithms: friends-of-friends (FOF), a percolation algorithm,
and DENMAX. (See below
for a description of how these algorithms work.) Fig.\ \ref{pfigure} reveals
that the two algorithms give very different results for the number of
massive halos in large simulations. In particular, DENMAX overpredicts
their number when compared to FOF or the theoretical prediction based upon
Press-Schechter theory. (Again, see below for an explanation of
Press-Schechter theory.) Thus the choice of group identification algorithm
indeed has an effect on the properties of halos, and this article attempts
to explore this by comparing three of the more popular algorithms.

Section 2 gives a brief overview of the N-body simulations
used to provide the `raw material' for the comparison. Section 3
introduces the three different algorithms which had been chosen for that
purpose. Section 4 describes Press-Schechter theory, which is used as
a baseline to compare the mass distribution of halos as identified by the
different algorithms, and how the agreement (or disagreement) with
Press-Schechter theory can be quantified. The next section contains the
actual results of the comparisons, how these depend upon the choice of
the free parameters in the algorithms, the box sizes of the simulation,
and the cosmology. This is then followed by the conclusions.

\section{The Simulations}

\begin{table*}
\begin{tabular}{||lr||c|c|c|c|c|c|r||} \hline\hline
\multicolumn{2}{||l||}{Simulation} & $\mathnormal{\Omega}_0$ & $h$ & $\sigma_8$ &
\# of & soft. & $z_{\mathrm{initial}}$ &
\multicolumn{1}{c||}{\# of} \\
& & & & & part. & length & &
\multicolumn{1}{c||}{steps} \\ \hline\hline
OCDM, & 50 Mpc & 0.4 & 0.6 & 0.8 & $64^3$ & 0.1 & 199 & 2022 \\
     & 100 Mpc & 0.4 & 0.6 & 0.8 & $64^3$ & 0.1 & 199 & 1333 \\
     & 200 Mpc & 0.4 & 0.6 & 0.8 & $64^3$ & 0.1 & 199 & 681 \\
     & 400 Mpc & 0.4 & 0.6 & 0.8 & $64^3$ & 0.1 & 199 & 337 \\
     & 800 Mpc & 0.4 & 0.6 & 0.8 & $64^3$ & 0.1 & 199 & 163 \\
    & 1600 Mpc & 0.4 & 0.6 & 0.8 & $64^3$ & 0.1 & 199 & 92 \\ \hline\hline
SCDM, & 50 Mpc & 1.0 & 0.5 & 1.0 & $64^3$ & 0.1 & 199 & 2865 \\
     & 100 Mpc & 1.0 & 0.5 & 1.0 & $64^3$ & 0.1 & 199 & 1880 \\
     & 200 Mpc & 1.0 & 0.5 & 1.0 & $64^3$ & 0.1 & 199 & 999 \\
     & 400 Mpc & 1.0 & 0.5 & 1.0 & $64^3$ & 0.1 & 199 & 459 \\
     & 800 Mpc & 1.0 & 0.5 & 1.0 & $64^3$ & 0.1 & 199 & 215 \\
    & 1600 Mpc & 1.0 & 0.5 & 1.0 & $64^3$ & 0.1 & 199 & 104 \\ \hline\hline
\end{tabular}
\caption{Parameters of the OCDM and SCDM simulations with varying
box sizes used in comparing the different group identification
algorithms. $h$ is the Hubble constant in units of 100 km/s/Mpc.
The gravitational softening length is given as a
fraction of the mean particle separation, and is kept constant in
comoving coordinates during the simulation. The value corresponds
to that of an equivalent Plummer softening length, although the Hydra code
uses a slightly different shape \cite{hydra}. Timestepping is
regulated automatically by the code, and the number of steps,
which may not be of equal length, necessary to reach redshift zero
is given in the last column.}
\label{simtable}
\end{table*}
To test the reliability of group identification algorithms we use
representative results of N-body simulations. The
first indications of the effects of different group
identification algorithms were found in N-body simulations with
the P$^3$M code by E. Bertschinger and J. M. Gelb \cite{bert}.
We decided to use a different code for this systematic study.
The simulations were run with the Hydra code by H. M. P. Couchman
et al.\ \cite{hydra}. Hydra incorporates the evolution of both,
dark matter via an adaptive particle-particle/particle-mesh
method (AP$^3$M), and of baryonic gas with smoothed particle hydrodynamics
(SPH). It was run with dark matter particles only since the goal was
to provide typical cosmological settings for halo identification.
Hydra was chosen because of its higher computational efficiency
in high density regions, which are the regions we are particularly
interested in, and because software to
identify groups in the output files produced by Hydra is readily available.
Two ``favorite'' cosmological scenarios were chosen: an
open cold dark matter model (OCDM) with $\mathnormal{\Omega}_0 = 0.4$, a Hubble
constant of $H_0 = 60 \mbox{ km/s/Mpc}$, normalized in such a way
that $\sigma_8 = 0.8$ at present time, consistent with the observed
abundance of galaxy clusters \cite{pen,eke}. The second model was
standard cold dark matter (SCDM) with $\mathnormal{\Omega}_0 = 1$,
$H_0 = 50 \mbox{ km/s/Mpc}$, and $\sigma_8 = 1$.
For each model, simulations were run with
$64^3$ particles with box sizes varying from 50 Mpc to 1600 Mpc in
steps of a factor of two both to insure a wide dynamic range in the masses
of the dark matter halos and to see at which resolution the
group identification algorithms break down. Table \ref{simtable} lists
all the parameters of the simulations, including the gravitational
softening employed, initial redshift, and the number of timesteps.

\section{Identifying Groups}

At redshift zero, groups in the distribution of dark matter particles
were identified using three common algorithms: the friends-of-friends (FOF)
method, DENMAX, and SKID\@. With FOF \cite{fof0,fof}, particles are joined into
groups if the separation to the nearest neighbor is less than a given
threshold, called the linking length $b$, which is the only free parameter
for this algorithm. We will express $b$ in units of the mean particle
separation throughout this paper. Then $1/b^3$ corresponds to an overdensity,
and FOF approximately groups together particles which lie inside the
corresponding level surfaces.

Group identification in DENMAX \cite{bert,DENMAX} and SKID \cite{SKID,SKID2}
is a three-step process. First, particles are moved in small steps in
the direction of the local density gradient, until their locations stay
within a small distance, the convergence radius, for a certain number of
steps, which means that they oscillate around a local density maximum.
In the second step, groups among the particles at their moved positions
are identified with a FOF algorithm with a linking length $b$ twice the
convergence radius. The third and last step then removes particles which
are not gravitationally bound to the groups identified in the previous step.
This is done by calculating the potential energies of all of the particles
in the group (now at their original, unmoved locations) and their kinetic
energies (with peculiar velocities with respect to the center of mass of
the group and an additional velocity due to the Hubble expansion of the
halo). If there are unbound particles with positive total energy,
the one with the highest total energy is removed from the group, and the
process of calculating potential and kinetic energies is repeated, until
there are no unbound particles left. Note that this implementation
of DENMAX actually follows SKID and is slightly different
from the originally proposed \cite{bert,DENMAX}. The difference between
DENMAX and SKID lies in the way densities, and thus their gradients,
are determined in the first step. For DENMAX, they were calculated on a
fixed grid with a triangular shaped cloud (TSC) scheme \cite[Ch. 5-3-2]{HE},
the same as in the P$^3$M code. SKID instead uses a symmetric SPH smoothing
kernel \cite{sph} considering a certain number of neighbors around each
particle to smooth over. Hence the main difference between the two
algorithms is that DENMAX computes density gradients on a fixed
Eulerian scale, whereas SKID computes them on a fixed Lagrangian scale,
adjusting the resolution of this calculation to the local density.
Different to FOF, DENMAX and SKID have two free parameters each:
the linking length $b$ and the size of the grid (for DENMAX) or the number of
nearest neighbors (for SKID) used in calculating the density gradients.

\section{Press-Schechter Theory and How to Compare With It}

One of the most basic quantities derived from identifying halos is the
distribution of their masses, which should be insensitive to different
algorithms. To this end, the differential mass function $n(M)$, i.e.\ the
number density of halos with masses between $M$ and $M+\rd M$ as obtained
in the simulations and by the different algorithms, was compared with the
prediction from Press-Schechter theory \cite{ps,bcek} based upon the
power spectrum $P(k)$ used for the initial conditions in the simulations
\be
\begin{array}{c}
n_\PS (M)\; \rd M = \\ -\sqrt{\frac{2}{\pi}}\frac{\bar{\rho}}{M^2}
\frac{\delta_\rc}{D\sigma(M)}\frac{\rd \ln \sigma}{\rd \ln M}
\exp\left[-\frac{\delta_\rc^2}{2D^2\sigma^2(M)}\right]\rd M.
\end{array}
\label{nps}
\ee
We use the form of the differential mass function as given by
\cite{lc} for redshift $z=0$, the case we are interested in.
$\bar{\rho}$ is the present mean mass density of the universe, $\delta_\rc$
the linearly extrapolated overdensity at which a top-hat shaped overdensity
has collapsed ($\delta_\rc=1.686$ for SCDM and $\delta_\rc=1.660$
for OCDM at $z=0$ --- see \cite{deltac} for a discussion of this quantity
and its dependence upon cosmological parameters and redshift). $\sigma$
is the fractional r.m.s. mass fluctuation within a top-hat of radius $R$ in
the initial conditions of the simulation
\[
\sigma^2(R)=4\pi\int_0^\infty k^2\rd k\; P(k) W_R^2(k),
\]
where $W_R(k)$ is the Fourier transform of the top-hat window function
\[
W_R(k)=3\left[\frac{\sin kR}{(kR)^3}-\frac{\cos kR}{(kR)^2}\right].
\]
The connection between $\sigma$ as a function of radius and sigma as a
function of mass, which is needed in evaluating eq.\ (\ref{nps}), is then
\[
M=\frac{4\pi}{3}R^3\bar{\rho},
\]
via the average mass
contained within the top hat. Finally, $D$ in eq.\ (\ref{nps}) is the ratio
of the growing modes of linear perturbation theory $D(z)$ at present time
and at the redshift of the initial conditions $D=D(z=0)/D(z=z_{\mathrm{initial}})$.
\begin{figure*}
\epsfxsize=5.865 in \epsfbox{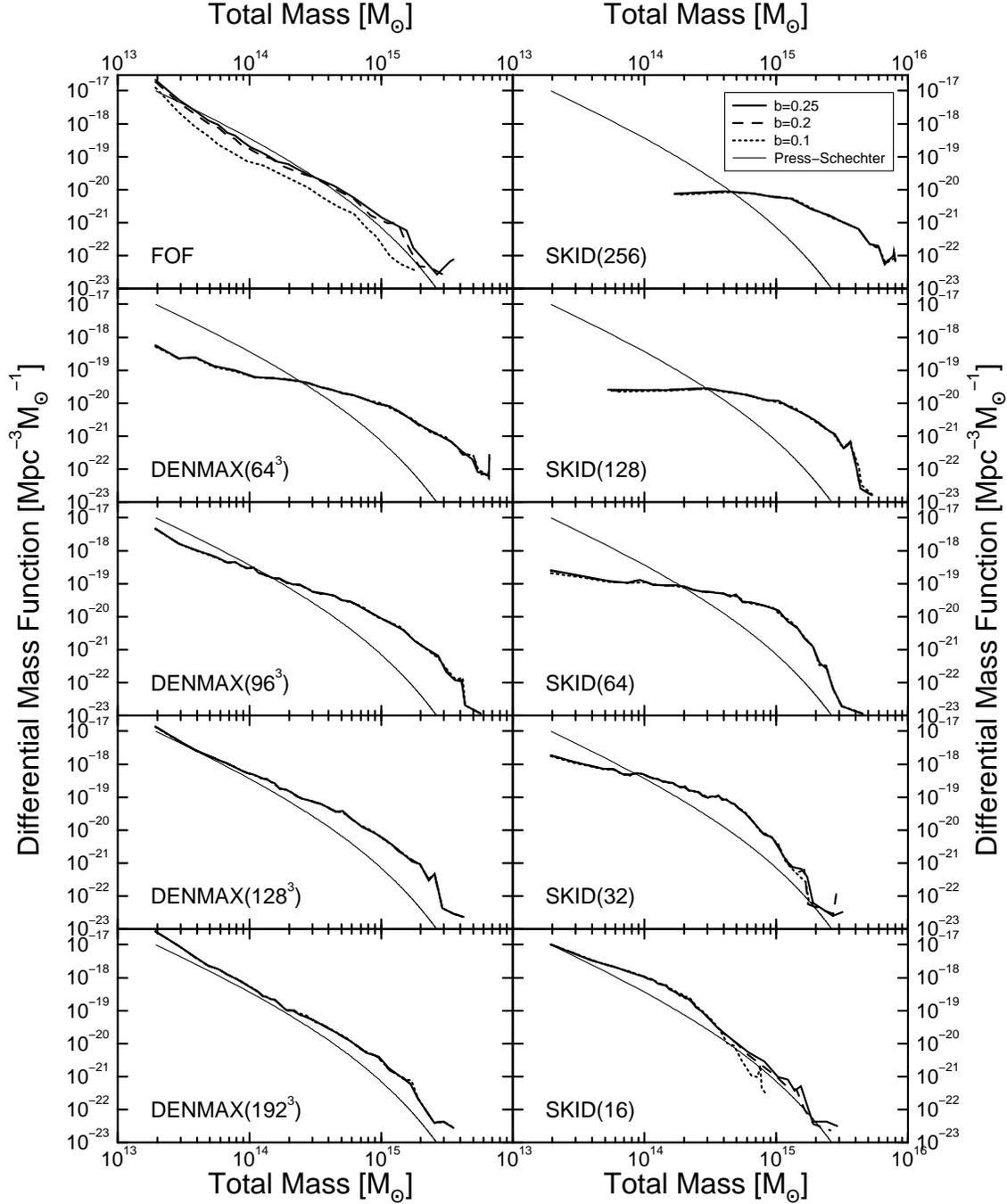}
\caption{The differential mass functions obtained by different group
finding algorithms and how they vary with their parameters. The halos
were all identified at redshift $z=0$ of the same OCDM simulation
($\mathnormal{\Omega}_0=0.4$, $h=0.6$) run with Hydra using $64^3$ dark matter
particles in a 400 Mpc box. The linking length $b$, a variable in
all of the three algorithms, is given in units of the mean particle
separation. In addition to $b$, DENMAX and SKID have as additional
parameters the grid size or the number of nearest neighbors used
to calculate the density gradients, respectively. These numbers are
given in parentheses. Note that for DENMAX and SKID the differential
mass function barely depends upon the linking length, contrary to
FOF\@. The differential mass function predicted from Press-Schechter
theory is plotted as a thin line in all of the graphs as a comparison.
The scatter at the high-mass ends of the curves is consistent with
Poisson noise from binning the halo masses in order to construct the
differential mass function.}
\label{bfigure}
\end{figure*}
\begin{table*}
\begin{tabular}{||lr||c|c|c||c||} \hline\hline
\multicolumn{2}{||l||}{Algorithm} & \multicolumn{3}{c||}{$b =$} & Range of \\
& & 0.25 & 0.2 & 0.1 & $\lg (M/M_\odot)$ \\ \hline\hline
FOF & & 0.110 & 0.098 & 0.505 & 13.3 $\ldots$ 15.25 \\ \hline\hline
DENMAX,& $64^3$ grid & 1.551 & 1.565 & 1.579 & 13.3 $\ldots$ 15.25 \\
& $96^3$ grid & 0.840 & 0.841 & 0.852 & 13.3 $\ldots$ 15.25 \\
& $128^3$ grid & 0.583 & 0.584 & 0.591 & 13.3 $\ldots$ 15.25 \\
& $192^3$ grid & 0.287 & 0.288 & 0.295 & 13.3 $\ldots$ 15.25 \\ \hline\hline
SKID,& 256 neighb.  & 0.641 & 0.647 & 0.663 & 14.3 $\ldots$ 15.25 \\
& 128 neighb.  & 1.379 & 1.389 & 1.416 & 13.8 $\ldots$ 15.25 \\
& 64 neighb.  & 1.838 & 1.862 & 1.921 & 13.3 $\ldots$ 15.25 \\
& 32 neighb.  & 0.633 & 0.624 & 0.620 & 13.3 $\ldots$ 15.25 \\
& 16 neighb.  & 0.221 & 0.193 & 0.213 & 13.3 $\ldots$ 15.25 \\ \hline\hline
\end{tabular}
\caption{Goodness of agreement $\mathnormal{\Delta}$ between the differential
mass function of halos as determined by several different algorithms and
its prediction from Press-Schechter theory, for the simulation
and set of parameters which are shown in Fig.\ \ref{bfigure}. The upper
and lower limits of the mass range, over which $\mathnormal{\Delta}$ is calculated,
are given in the last column. The numbers for SKID with 256 and 128
neighbors are smaller only because the mass range had to be
restricted as no small groups were identified in these two cases.
Furthermore, because of lack of high-mass halos, $\mathnormal{\Delta}$
with SKID with 16 nearest neighbors and $b=0.1$ was calculated for
$\lg (M/M_\odot) = 13.3 \dots 14.9$.}
\label{btable}
\end{table*}

In the interpretation and derivation of \cite{bcek,lc}, the
Press-Schechter prediction (\ref{nps}) gives the mass distribution of
regions which lie inside contours with density contrast $\delta_\rc$,
which is the linearly extrapolated density contrast at which a full
calculation, which is possible for a top-hat shaped overdensity,
shows that the region has already collapsed.

Since one of the group
identification algorithms (FOF) showed a good agreement with
Press-Schechter theory in the simulations which drew our attention
to this problem (Fig.\ \ref{pfigure}), we decided to use Press-Schechter
theory as a baseline for comparing the different algorithms.
To quantify the differences in the differential mass functions obtained
in the simulations, $n(M)$, and from Press-Schechter theory, $n_\PS(M)$, we
simply integrate the difference squared between appropriate masses
$M_0$ and $M_1$ to obtain a goodness of agreement
$\mathnormal{\Delta}$
\be
\mathnormal{\Delta}=\int_{\lg M_0}^{\lg M_1} \left[\lg n(M) - \lg n_\PS(M)\right]^2
\;\rd\lg M,
\label{Delta}
\ee
but with the logarithms taken of all relevant quantities. (Here, we
use logarithms to base 10 and express the differential mass functions
in units of Mpc$^{-3}M_\odot^{-1}$ and masses in $M_\odot$.) Since the
differential mass functions approximately drop as a power law with
increasing mass, introducing the logarithms in (\ref{Delta}) assures
that differences at the low-mass end contribute as much to
$\mathnormal{\Delta}$ as differences at the high-mass end.

\section{Results}

The three different group finding algorithms, with varying parameters,
were applied to the results of the N-body simulations, and the resulting
differential mass functions of halos were compared with
Press-Schechter theory. First, we looked at how the parameters
of the algorithms influence the mass function, and then how varying the
box size of the simulation (and thus effectively changing the resolution
of the halos) affected the results.

\subsection{Varying the Parameters}

Fig.\ \ref{bfigure} represents the differential mass functions in a typical
case, here a 400 Mpc box with $64^3$ particles from the series of OCDM
Hydra simulations. The linking lengths used are $b=0.25$, $b=0.2$,
and $b=0.1$, in units of the mean particle separation. These values cover
the range used in the literature. We see that the differential mass
function obtained with FOF depends strongly upon $b$, and that the
value $b=0.2$ gives good agreement with Press-Schechter theory.
(This has been noticed earlier in \cite{efwd} and \cite{laceycole} ---
see \cite{virial} for a comparison of FOF with virialized halos.) For
FOF, the different values of the linking length effectively correspond
to different values of the density contours, inside which
particles are grouped together, and therefore we expect a strong dependence
upon that parameter. Since FOF does not require the calculation of particle
densities, $b$ is the only free parameter for this algorithm. In
contrast, DENMAX and SKID show hardly any variation with the linking
length. This is not surprising. During the first phase of these two
algorithms the particles are moved towards the closest density
maximum until their locations vary by less than the convergence radius.
But the linking length of the FOF step during the second phase is twice
the convergence radius. Hence changing $b$ also changes how tight the
moved particles cluster around the density maxima. But the location of
the density maxima is independent of the linking length, and so the FOF
step scoops up the same particles into the same groups before the removal
of gravitationally unbound particles takes place, which is based upon
the original, unmoved locations.

\begin{figure*}
\epsfxsize=5 in \epsfbox{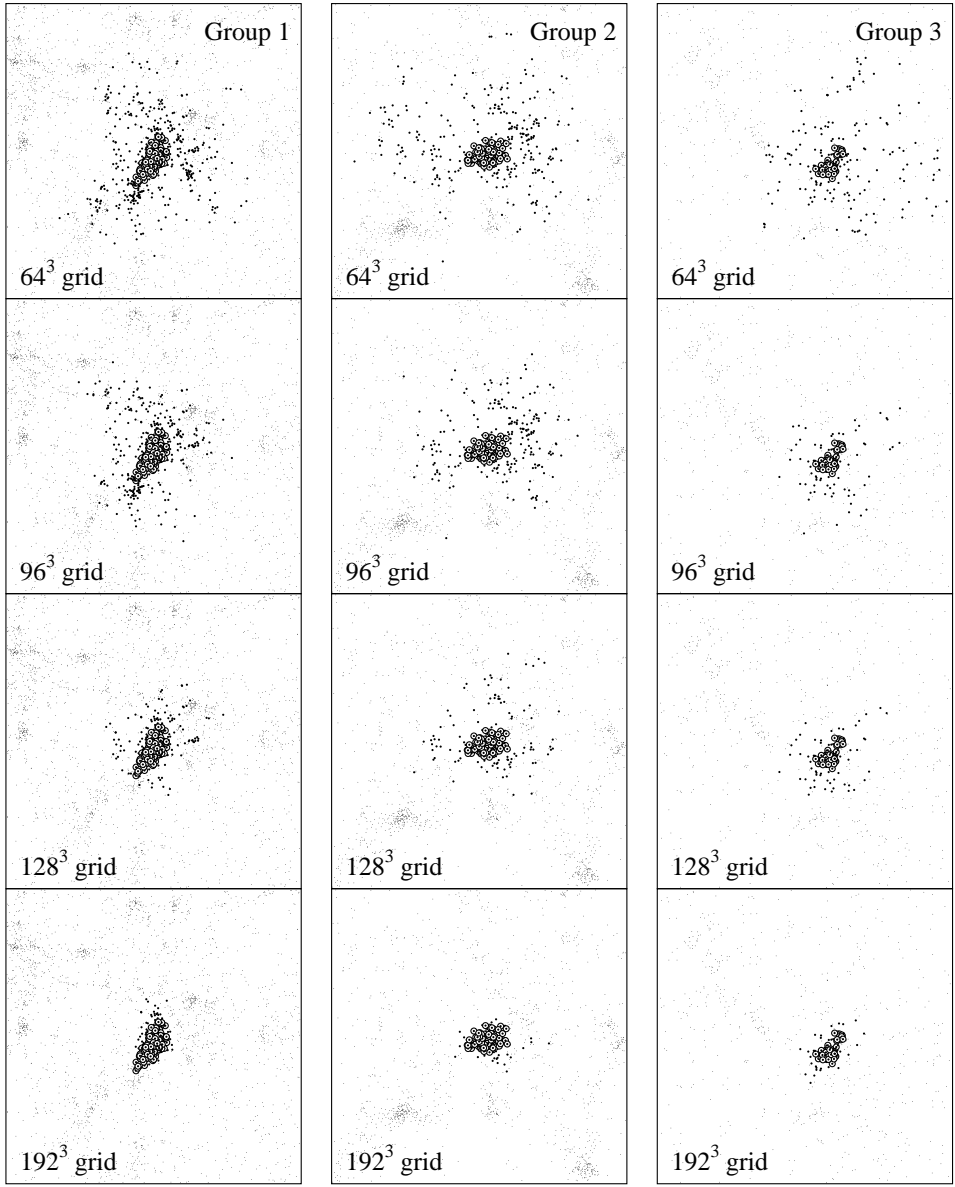}
\caption{Three representative groups from the 400 Mpc OCDM simulation,
as identified by DENMAX and FOF, both with linking length $b=0.2$, but
different grid resolution in the calculation of the density gradients
for DENMAX. Particles, which are members of the corresponding DENMAX
group, are marked by filled circles, while those in the FOF groups are
marked by additional open circles. Particles, which do not belong to the
DENMAX groups, are shown as dots. The left panel (for group 1) shows a
(70 Mpc)$^3$ cube, the central and right one (for groups 2 and 3) a
(50 Mpc)$^3$ cube, each projected along the $z$ axis of the simulation.}
\label{fdfigure}
\end{figure*}
\begin{figure*}
\epsfxsize=5 in \epsfbox{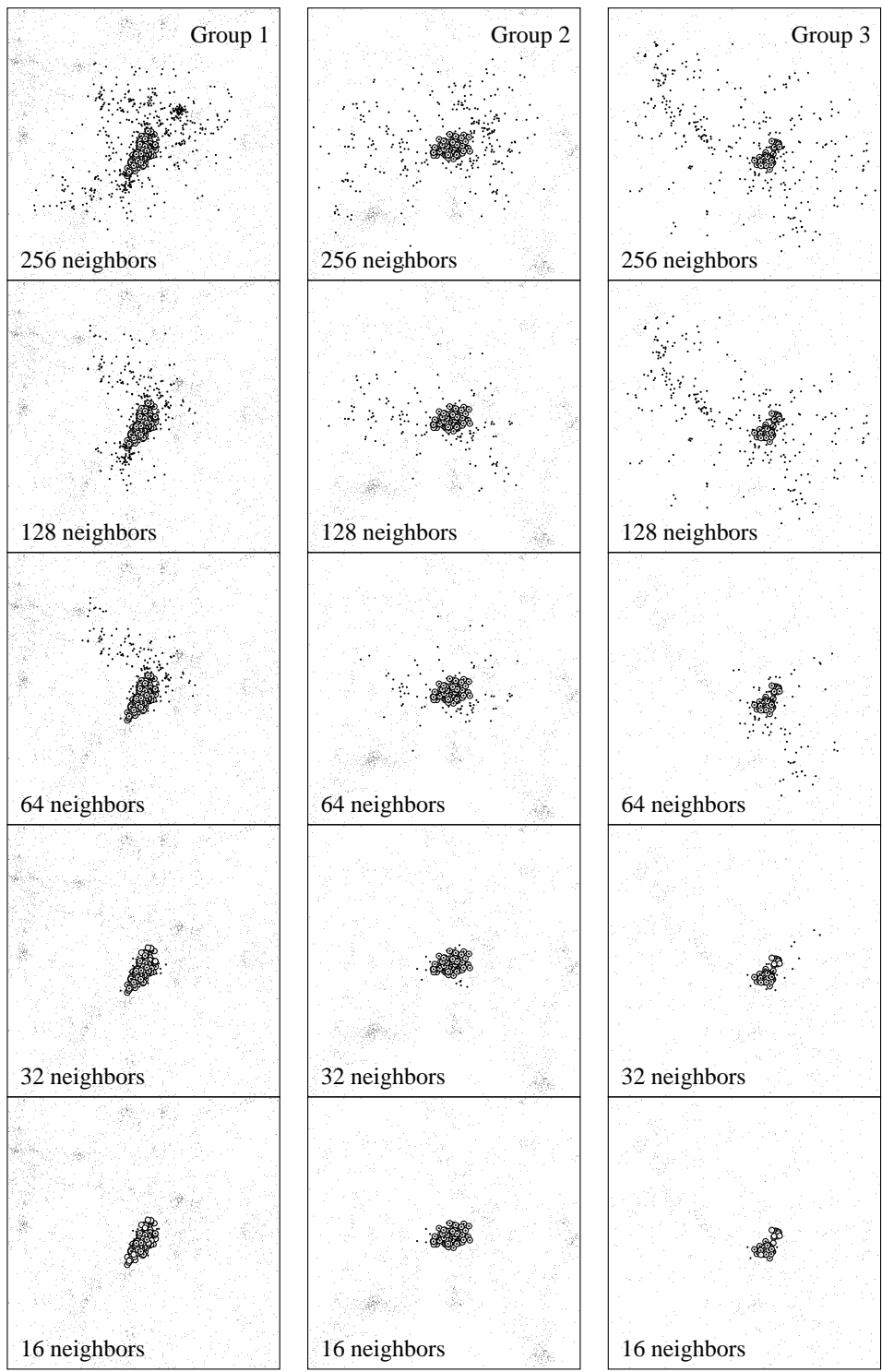}
\caption{The same three groups as in Fig.\ \ref{fdfigure}, but now
the filled circles show the groups as identified with SKID at different
resolutions of the density field (with $b=0.2$).}
\label{fsfigure}
\end{figure*}
\begin{figure*}
\epsfxsize=5.865 in \epsfbox{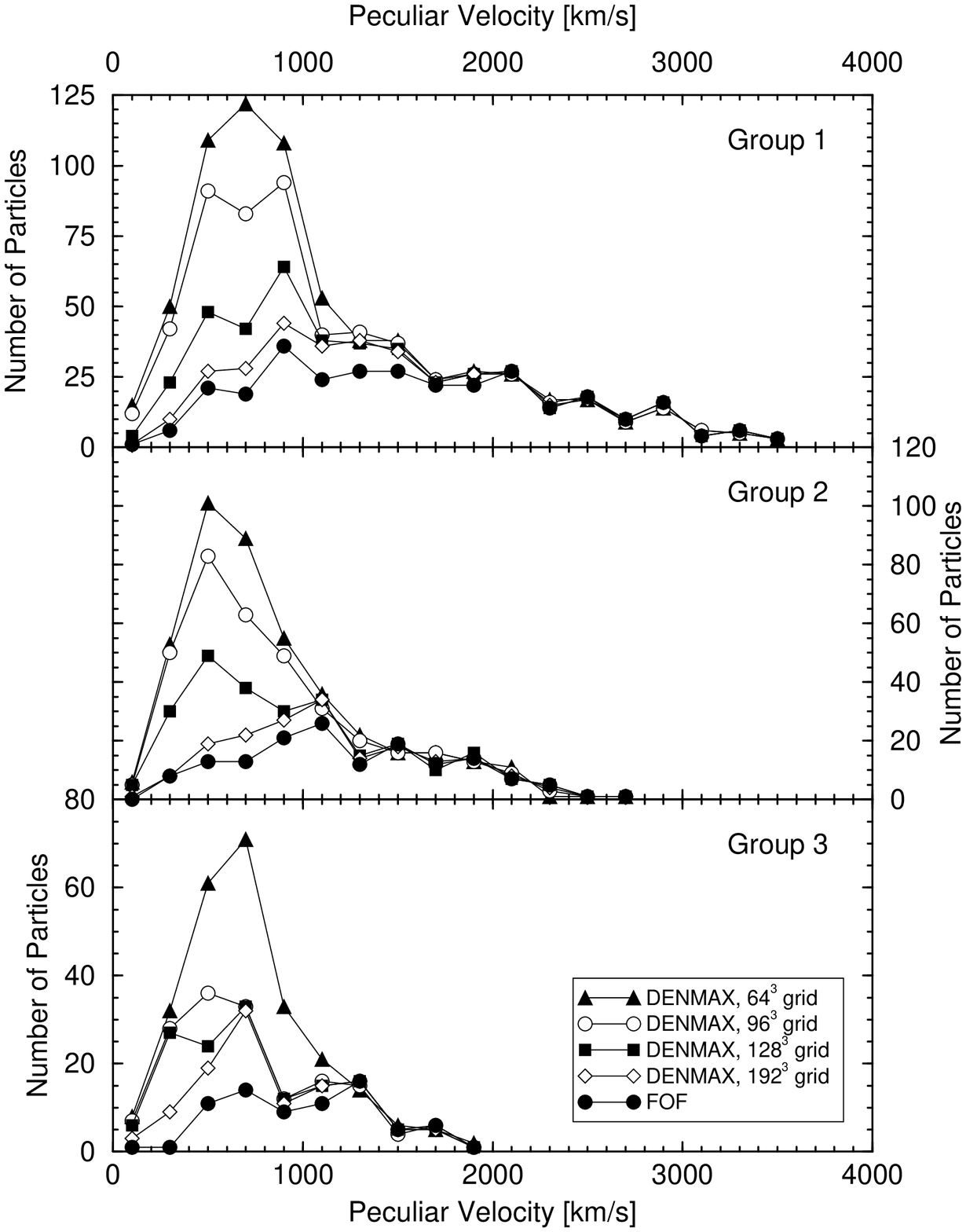}
\caption{The distribution of peculiar velocities with respect to the
center of mass in the three groups
from Fig.\ \ref{fdfigure} for FOF and the different resolutions of
DENMAX (with $b=0.2$). The peculiar velocities contain a component
from the Hubble expansion of the group.}
\label{vdfigure}
\end{figure*}
\begin{figure*}
\epsfxsize=5.865 in \epsfbox{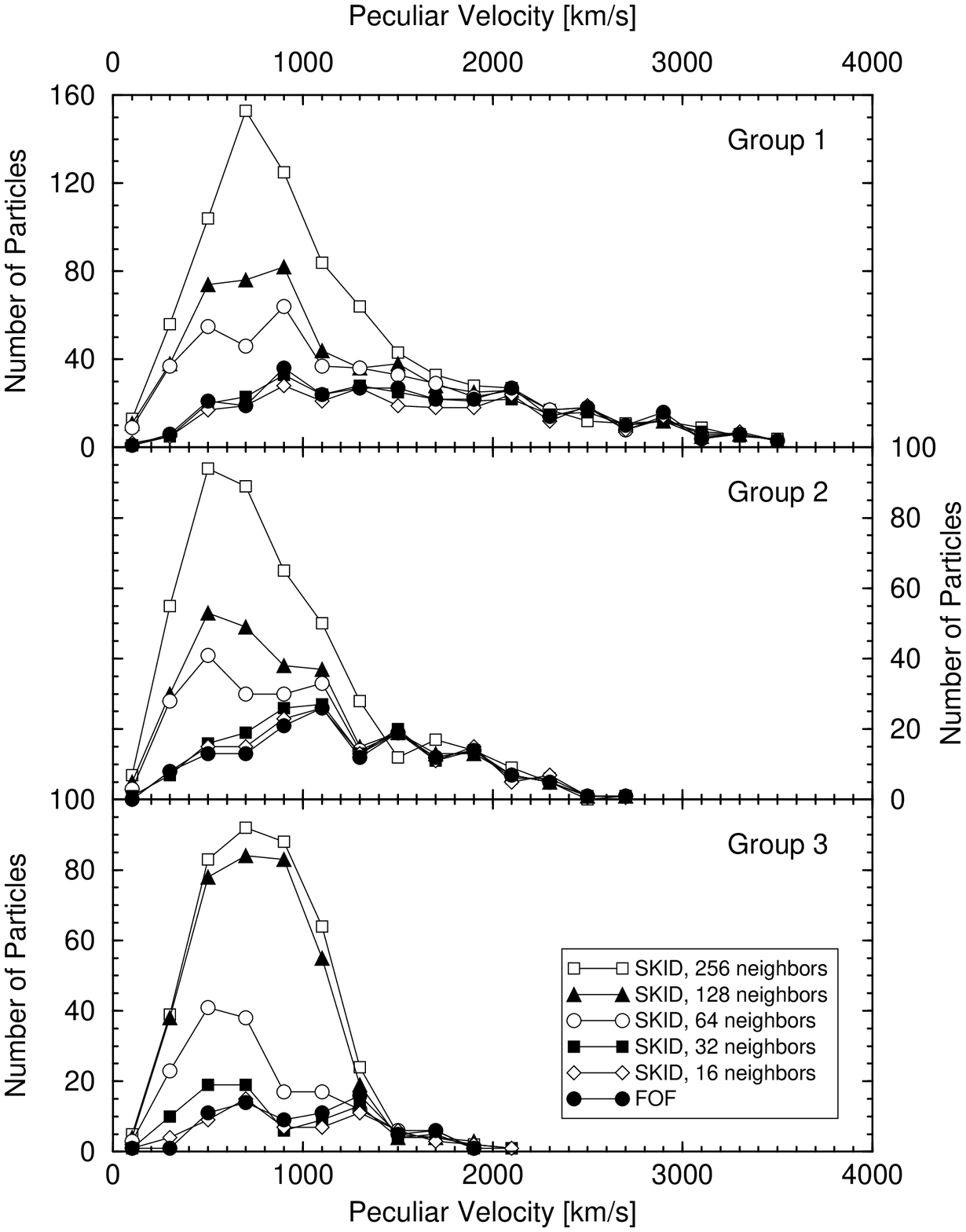}
\caption{The distribution of peculiar velocities with respect to the
center of mass in the three groups
from Fig.\ \ref{fsfigure} for FOF and the different resolutions of
SKID (with $b=0.2$). The peculiar velocities contain a component
from the Hubble expansion of the group.}
\label{vsfigure}
\end{figure*}
\begin{table*}
\begin{tabular}{||lr||r|r||r|r||r|r||} \hline\hline
\multicolumn{2}{||l||}{Algorithm} & \multicolumn{2}{c||}{Group 1} &
\multicolumn{2}{c||}{Group 2} & \multicolumn{2}{c||}{Group 3} \\
& & \multicolumn{1}{c|}{\# of} & \multicolumn{1}{c||}{$\sigma_{\mathrm{1D}}$}
& \multicolumn{1}{c|}{\# of} & \multicolumn{1}{c||}{$\sigma_{\mathrm{1D}}$}
& \multicolumn{1}{c|}{\# of} & \multicolumn{1}{c||}{$\sigma_{\mathrm{1D}}$} \\
& & \multicolumn{1}{c|}{part.} & \multicolumn{1}{c||}{[km/s]}
& \multicolumn{1}{c|}{part.} & \multicolumn{1}{c||}{[km/s]}
& \multicolumn{1}{c|}{part.} & \multicolumn{1}{c||}{[km/s]} \\ \hline\hline
FOF & & 303 & 1037 & 152 & 781 & 75 & 626 \\ \hline\hline
DENMAX, & $64^3$ grid & 681 & 768 & 421 & 561 & 253 & 462 \\
        & $96^3$ grid & 587 & 811 & 360 & 579 & 158 & 486 \\
       & $128^3$ grid & 438 & 910 & 260 & 642 & 144 & 502 \\
       & $192^3$ grid & 365 & 981 & 184 & 739 & 116 & 548 \\ \hline\hline
SKID,  & 256 neighb.  & 800 & 753 & 446 & 566 & 407 & 489 \\
       & 128 neighb.  & 549 & 834 & 286 & 629 & 372 & 485 \\
        & 64 neighb.  & 466 & 882 & 238 & 663 & 164 & 491 \\
        & 32 neighb. & 293 & 1031 & 168 & 757 &  88 & 563 \\
        & 16 neighb. & 265 & 1052 & 159 & 770 &  66 & 620 \\ \hline\hline
\end{tabular}
\caption{The number of particles and the 1D velocity dispersions
$\sigma_{\mathrm{1D}}$ (averaged over all possible lines of sights) for the
three groups in Figs.\ \ref{fdfigure} through \ref{vsfigure} as
identified by the different group finding algorithms (all have
$b=0.2$).}
\label{grouptable}
\end{table*}
Instead, DENMAX and SKID show a strong dependence upon the resolution at
which the density gradients are calculated. (The resolution of the
simulation stays the same throughout this discussion as the number of
particles is kept the same.) For DENMAX, the grid size
was increased from $64^3$ (the same as the number of particles) to
$192^3$ (the largest size handled by the workstation on which the calculations
were performed). For SKID, the number of nearest neighbors was reduced
from 256 to 16. It can be seen from Fig.\ \ref{bfigure} that a low
resolution of the density field (corresponding to a small grid or
a large number of neighbors) produces preferably large halos, since there are
few density maxima present inside the volume towards which all of the
particles are moved. Each of the density maxima will then end up with
a large number of particles, and hence larger groups are
produced. On the other hand, with high resolution of the density field
(i.e.\ large grids or small number of neighbors), it becomes
bumpy, and, except for the densest regions, each particle will correspond
to a density maximum at its location. Thus the particles will not move much
during the first phase of DENMAX and SKID, and these algorithms become
close to plain FOF. This can be seen clearly
in Fig.\ \ref{bfigure} where the curves for DENMAX and SKID come closer
to the Press-Schechter prediction (to which FOF with $b=0.2$ is very close
as seen above) as the resolution of the density field increases.
We also see an emergence
of variation in the differential mass function with the linking length
for the high resolution version of SKID with 16 neighbors, as we would
expect if this algorithm became more similar to FOF\@.

Table \ref{btable} summarizes these results in numeric form. FOF with
$b=0.2$ has the overall smallest value of the goodness of agreement
$\mathnormal{\Delta}$. This is the case even when compared to those values for SKID,
where a smaller mass range for $\mathnormal{\Delta}$ had to be used because of the lack
of groups at the low-mass or high-mass ends. The values of $\mathnormal{\Delta}$ in
Table \ref{btable} for SKID with 256, 128, and 16 neighbors (there for
$b=0.1$ only) are only lower limits because of the restricted mass
ranges. For DENMAX and SKID we see, at the same grid size
or number of neighbors, very little variation of $\mathnormal{\Delta}$ with the
linking length, but an improvement towards the values obtained with
FOF as the resolution is increased.

To see more clearly where the large halos in low-resolution DENMAX
and SKID originate we have isolated three typical halos in the
400 Mpc OCDM simulation, confining ourselves to a linking length of
$b=0.2$ for the moment. Figs.\ \ref{fdfigure} and \ref{fsfigure} compare
the group identifications of FOF with those of DENMAX and SKID,
respectively. We see that low-resolution DENMAX and SKID indeed produce
very extended halos, as argued above from the smaller number of
density maxima in the simulation volume. The gravitational unbinding
step in both algorithms does not seem to be able get rid of all of these
outlying particles, which based upon the visual appearance should not be
regarded as members of the halo. As can be seen from Figs.\
\ref{vdfigure} and \ref{vsfigure}, it retains particles which have
a low peculiar velocity with respect to the center of mass of the halo
in these outlying regions. There is no shortage of such particles,
as we expect the particles in these regions to have low peculiar
velocities due to the shallow potential there. The situation improves
for the higher resolution versions of DENMAX and SKID when these outlying
particles are not considered to be group members in the first place
because these two algorithms become closer to plain FOF.

The inclusion of outlying particles with low peculiar velocities in
low-resolution DENMAX and SKID biases the velocity dispersion to lower
values in these cases, as can be seen in Table \ref{grouptable}. Thus
the choice of group identification not only affects the masses of the
halos, but also actual physical observables like the 1D velocity
dispersion of a halo.

The same results are observed in the 400 Mpc box of the SCDM simulation.

\subsection{Varying the Box Size}

\begin{figure*}
\epsfxsize=6 in \epsfbox{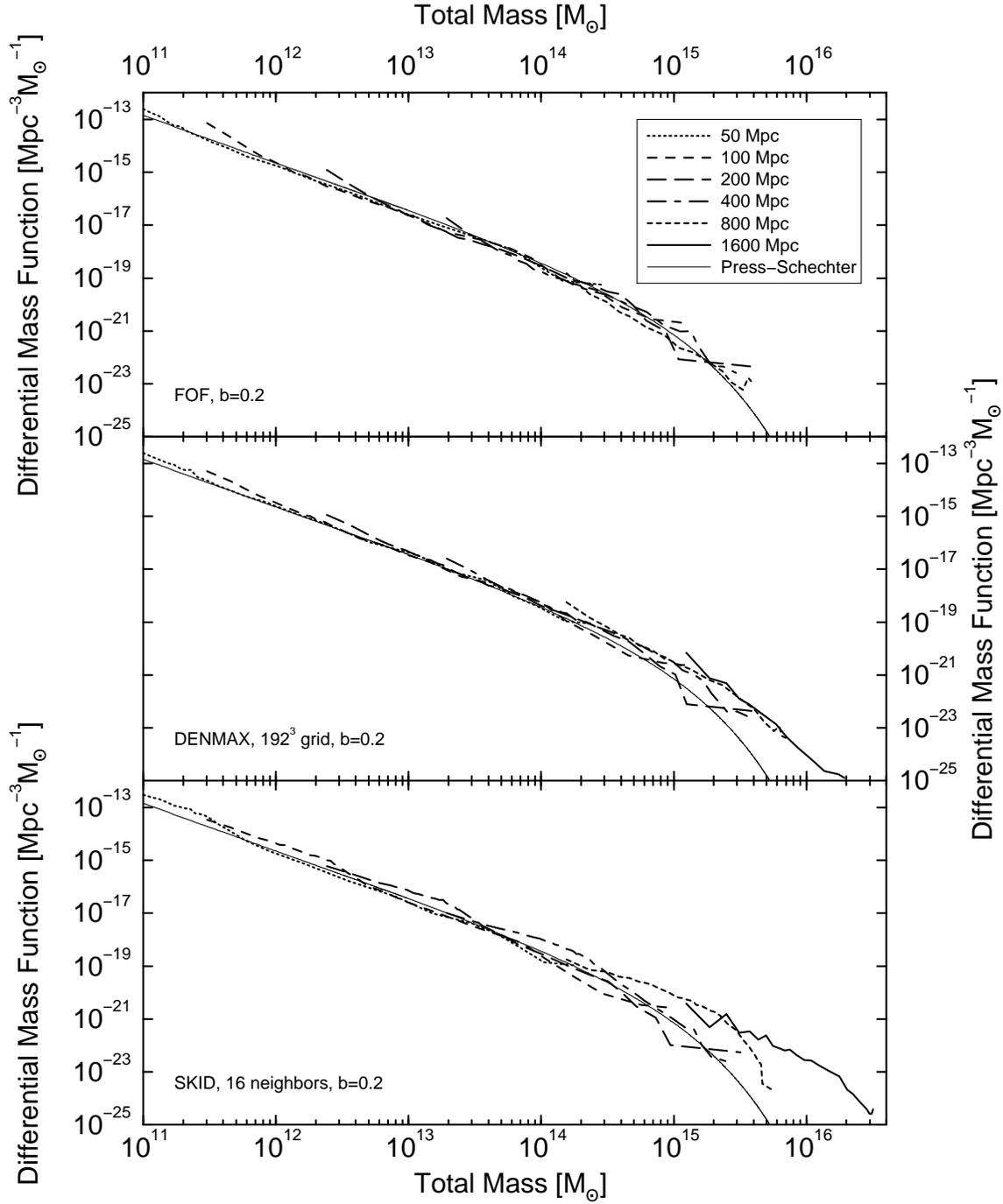}
\caption{The differential mass functions at redshift $z=0$ for the
OCDM Hydra simulation ($\mathnormal{\Omega}_0=0.4$, $h=0.6$) with $64^3$ dark matter
particles and box sizes varying from 50 Mpc to 1600 Mpc. FOF was run
with linking length $b=0.2$ (in units of the mean particle separation),
and DENMAX and SKID with the same $b$ and the highest resolution for
calculating the density field, i.e.\ a $192^3$ grid or 16 neighbors,
respectively. For the 1600 Mpc box, FOF produced too few groups to
successfully construct a reliable differential mass function, which is
thus missing from the plot. The scatter at the high-mass end is consistent
with Poisson noise.}
\label{ofigure}
\end{figure*}
\begin{figure*}
\epsfxsize=6 in \epsfbox{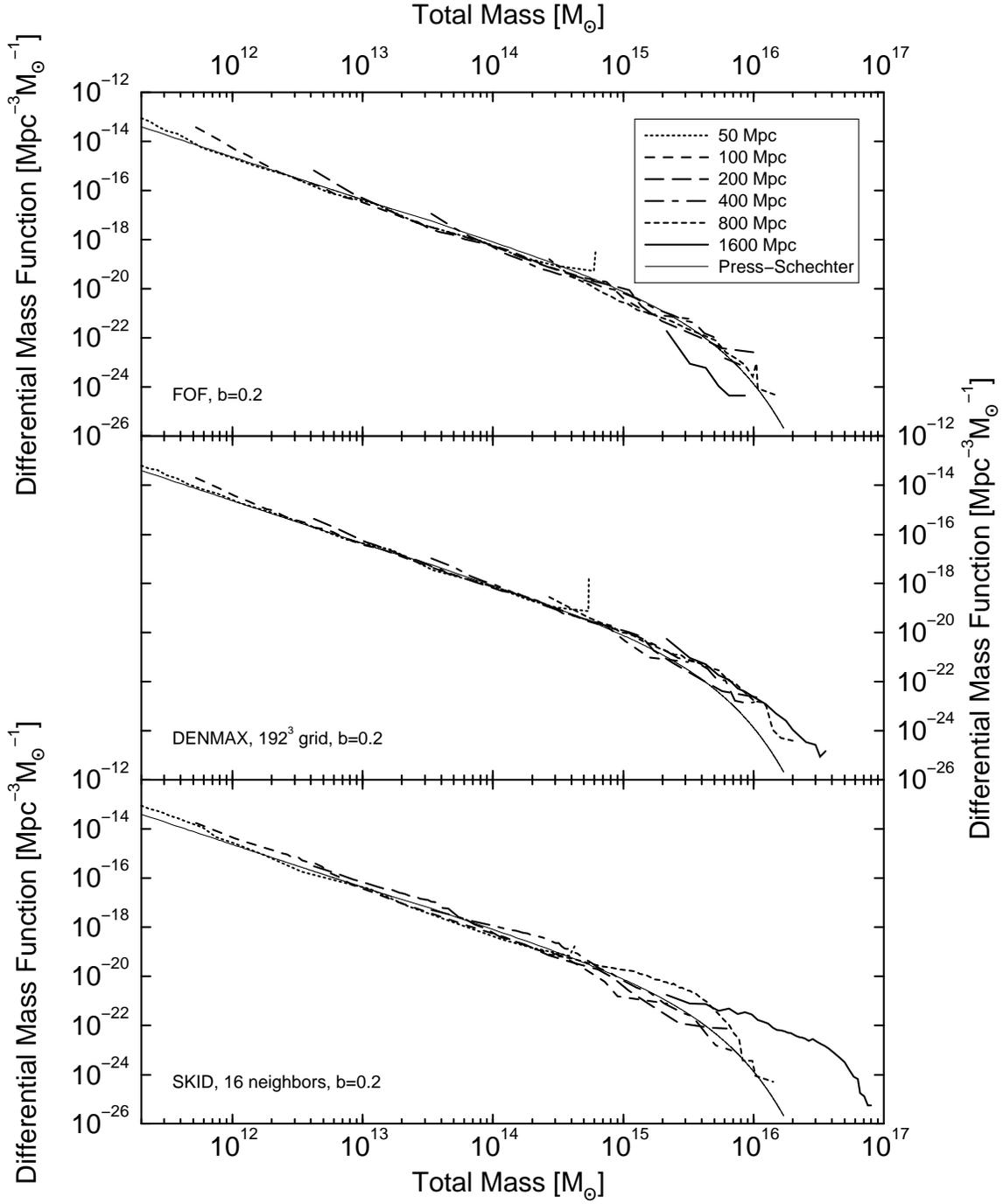}
\caption{The same as Fig.\ \ref{ofigure}, but now for the SCDM Hydra
simulations ($\mathnormal{\Omega}_0=1$, $h=0.5$) with $64^3$ dark matter
particles. With the 1600 Mpc box, FOF again produced very few groups.
Although a nominal differential mass function was determined and is
shown in this plot, it is not reliable and will not be used in the
further discussions.}
\label{sfigure}
\end{figure*}
\begin{table*}
\begin{tabular}{||l||c|c|c|c|c||} \hline\hline
Algorithm & \multicolumn{5}{c||}{Box size in Mpc} \\
&   50  &  100  &  200  &  400  &  800 \\ \hline\hline
FOF, $b=0.2$              & 0.038 & 0.068 & 0.149 & 0.101 & 0.063 \\
DENMAX, $192^3$ grid, $b=0.2$ & 0.023 & 0.037 & 0.094 & 0.307 & 0.494 \\
SKID, 16 neighbors, $b=0.2$ & 0.118 & 0.186 & 0.238 & 0.191 & 1.119 \\
\hline\hline
Range of           &  11.0 &  11.9 &  12.8 &  13.7 &  14.6 \\
$\lg (M/M_\odot)$  &  14.0 &  14.9 &  15.4 &  15.4 &  15.4 \\ \hline\hline
\end{tabular}
\caption{Goodness of agreement $\mathnormal{\Delta}$ between the differential
mass function of halos as determined by several different algorithms and
its prediction from Press-Schechter theory, for the OCDM simulations
which are shown in Fig.\ \ref{ofigure}. The upper
and lower limits of the mass range, over which $\mathnormal{\Delta}$ is calculated,
are given in the last two rows.}
\label{otable}
\end{table*}
\begin{table*}
\begin{tabular}{||l||c|c|c|c|c||} \hline\hline
Algorithm & \multicolumn{5}{c||}{Box size in Mpc} \\
&   50  &  100  &  200  &  400  &  800 \\ \hline\hline
FOF, $b=0.2$                & 0.066 & 0.089 & 0.126 & 0.094 & 0.093 \\
DENMAX, $192^3$ grid, $b=0.2$ & 0.025 & 0.028 & 0.023 & 0.090 & 0.382 \\
SKID, 16 neighbors, $b=0.2$ & 0.126 & 0.245 & 0.214 & 0.081 & 0.573 \\
\hline\hline
Range of           &  11.3 &  12.2 &  13.1 &  14.0 &  14.9 \\
$\lg (M/M_\odot)$ &  14.3 &  15.2 &  15.7 &  15.7 &  16.0 \\ \hline\hline
\end{tabular}
\caption{Goodness of agreement $\mathnormal{\Delta}$ between the differential
mass function of halos as determined by several different algorithms and
its prediction from Press-Schechter theory, for the SCDM simulations
which are shown in Fig.\ \ref{sfigure}. The upper
and lower limits of the mass range, over which $\mathnormal{\Delta}$ is calculated,
are given in the last two rows.}
\label{stable}
\end{table*}
To compare the group finding algorithms at different resolutions of the
simulations we looked at different box sizes (since the number of dark
matter particles is kept the same). The best values for the parameters
found in the discussion above are used, i.e.\ $b=0.2$ for the linking length
of FOF and the highest resolution in calculating the density field for
DENMAX and SKID. Since the choice of $b$ for DENMAX and SKID does not make
much of a difference, the same value $b=0.2$ was chosen for these two
algorithms.

The resulting differential mass functions are plotted in Fig.\ \ref{ofigure}
for the OCDM Hydra simulations, and in Fig.\ \ref{sfigure} for the SCDM
Hydra simulations. At visual inspection, the three algorithms seem to
perform equally up to box sizes of 200 Mpc (for OCDM) and 400 Mpc (for
SCDM), above which DENMAX and SKID start again to produce a larger
number of massive halos as compared with FOF or Press-Schechter
theory. At the low-mass ends, FOF and DENMAX show an overabundance
with respect to Press-Schechter theory. But the smallest groups
plotted in Figs.\ \ref{ofigure} and \ref{sfigure} consist of only two
particles, and hence this overabundance is not worrisome as such small
groups would be neglected in real applications anyhow.

The goodness of agreement $\mathnormal{\Delta}$ with Press-Schechter theory, calculated
over a mass range which excludes these smallest groups, confirms the trend
in the plots. For the OCDM Hydra simulations, Table \ref{otable} shows that
actually DENMAX has the best agreement up to 200 Mpc, after which FOF
fares the best. SKID is always worse than FOF and, except for one case,
even DENMAX. Similarily, Table \ref{stable} for the SCDM Hydra simulations
prefers DENMAX up to 200 Mpc, SKID slightly for 400 Mpc, and FOF for
800 Mpc, when compared with Press-Schechter theory. For the 1600 Mpc
simulations, FOF produced a very small number
of halos, such that it was not possible to construct a reliable mass function
in the OCDM model, and only a very unreliable one in the SCDM model.
Therefore, the 1600 Mpc box has been omitted from the comparison in Tables
\ref{otable} and \ref{stable}.

\section{Conclusions}

The most basic property of a dark matter halo, its mass,
depends upon the choice of group finding algorithm and its free parameters.
Many of the other properties are likely to do so as well, and we have
shown circumstantial evidence of this for the velocity dispersion. Much care
has to be taken in choosing the right algorithm for the purpose in
question. If the task is to reproduce the mass distribution as predicted
by Press-Schechter theory (which might not be correct, though), the
popular DENMAX and SKID algorithms
have to be run with a very high resolution of the density field.
For DENMAX this means a large grid (larger than the number of particles),
and for SKID a small number of neighbors (smaller than 64 neighbors, which
is the default number recommended by the authors of SKID \cite{SKID}).
Under these circumstances, DENMAX and SKID become close to a simple
FOF algorithm, which, for the right linking length, generally gives the best
agreement with Press-Schechter theory and at a much smaller computational
cost. Similar effects are seen when instead of the resolution of the algorithm
the resolution of the simulation is changed. With fixed number of particles,
DENMAX and SKID perform worse, with respect to Press-Schechter theory,
as the box size is increased, with FOF faring better. These results hold
up for both cosmological scenarios studied.

In cases with low resolution, DENMAX and SKID produce more
massive halos as compared with Press-Schechter theory and FOF.
These halos are larger because DENMAX and SKID have the tendency of
including distant particles which have a low peculiar velocity, a
problem from which FOF does not seem to suffer. This would make FOF
the preferable algorithm, and imply with its close agreement with
Press-Schechter theory, that its predictions are doing quite well, too.

\section*{Acknowledgments}

We wish to thank Edmund Bertschinger for providing the
P$^3$M code, and Hugh Couchman and the Hydra Consortium and the
University of Washington group of the High Performance Computing and
Communication Program for making the N-body simulation and group
identification codes available to the cosmology community on the World
Wide Web. Further thanks go to Lars Hernquist, Ue-Li Pen, and
David Weinberg for discussions and comments on a first draft of this
paper. Part of the computational work in support of this research was
performed at the Theoretical Physics Computing Facility at Brown University.
MG is supported by an SAO Predoctoral Fellowship and a Manning Fellowship
from Brown University.

\end{document}